\documentclass[aps,preprint,amsmath,amssymb,amsfonts]{revtex4}
\usepackage{epsfig}
\usepackage{graphicx}
\usepackage{subfigure}
\usepackage{dcolumn}
\usepackage{bm}
\usepackage{amsthm}
\usepackage{enumitem}
\usepackage{slashed}
\usepackage{braket}
\usepackage{amsmath}
\usepackage{mathtools}
\usepackage{bbold}

\usepackage{color}   
\usepackage{hyperref}
\hypersetup{
	colorlinks=true,  
	linkcolor=black,  
	citecolor=black,
	filecolor=black,
	urlcolor=black,
}

\makeatletter
\def\l@subsubsection#1#2{}
\makeatother

\newcommand{\const}{\mbox{const}}
\newcommand{\del}{\partial}

\newcommand{\scri}{\mathcal{I}}

\newcommand{\dSCFT}[2]{dS\textsubscript{#1}/CFT\textsubscript{#2}}

\begin{document}

\title{Higher-spin self-dual General Relativity:\\6d and 4d pictures, covariant vs. lightcone}

\author{Yasha Neiman}
\email{yashula@icloud.com}
\affiliation{Okinawa Institute of Science and Technology, 1919-1 Tancha, Onna-son, Okinawa 904-0495, Japan}

\date{\today}

\begin{abstract}
We study the higher-spin extension of self-dual General Relativity (GR) with cosmological constant, proposed by Krasnov, Skvortsov and Tran. We show that this theory is actually a gauge-fixing of a 6d diffeomorphism-invariant Abelian theory, living on (4d spacetime)$\times$(2d spinor space) modulo a finite group. On the other hand, we point out that the theory respects the 4d geometry of a self-dual GR solution, with no backreaction from the higher-spin fields. We also present a lightcone ansatz that reduces the covariant fields to one scalar field for each helicity. The field equations governing these scalars have only cubic vertices. We compare our lightcone ansatz to Metsaev's lightcone formalism. We conclude with a new perspective on the lightcone formalism in (A)dS spacetime: not merely a complication of its Minkowski-space cousin, it has a built-in Lorentz covariance, and is closely related to Vasiliev's concept of unfolding.
\end{abstract}

\maketitle
\tableofcontents
\newpage

\section{Introduction} \label{sec:intro}

\subsection{Higher-spin gravity: promise and difficulties}

Higher-spin (HS) gravity \cite{Vasiliev:1990en,Vasiliev:1995dn,Vasiliev:1999ba} is the putative interacting theory of an infinite tower of massless gauge fields with increasing spin; in its minimal version, it has a single field of every even spin $s=0,2,4,\dots$. As such, it is simultaneously a larger version of supergravity, and a smaller version of string theory. It shares some of string theory's healthy features, such as an AdS/CFT holographic formulation \cite{Klebanov:2002ja,Sezgin:2002rt,Sezgin:2003pt,Giombi:2012ms}, and a relationship between BPS solutions and fundamental objects \cite{David:2020fea,Lysov:2022zlw}. However, unlike string theory, it is ``native'' to 4 spacetime dimensions, and is equally happy with either sign of the cosmological constant: in particular, its AdS/CFT duality can be extended to $\Lambda>0$, providing a working model of \dSCFT{4}{3} \cite{Anninos:2011ui}. The grain of salt here is that, in its simplest version (which is the one compatible with dS/CFT), HS theory does not have a limit where the higher-spin fields decouple, or where the graviton's interactions are those of GR. 

The original formulation of HS theory, via the Vasiliev equations, is very indirect, and isn't manifestly local (its actual degree of non-locality is a subject of much work and debate \cite{Fotopoulos:2010ay,Taronna:2011kt,Vasiliev:2015wma,Bekaert:2015tva,Skvortsov:2015lja,Sleight:2017pcz,Ponomarev:2017qab,Gelfond:2018vmi,Didenko:2018fgx,Didenko:2019xzz,Gelfond:2019tac,Vasiliev:2022med,Neiman:2023orj}; the optimistic expectation is non-locality within a \emph{cosmological radius}, while the pessimistic one is non-locality at \emph{all length scales}; the author is a cautious optimist). The formalism is quite interesting in its own right. Here, we will highlight two of its aspects. The first is the use of twistor coordinates alongside the usual spacetime ones. The second is the use of \emph{master fields} -- a higher-spin version of superfields, which carry the fields of all spins \emph{together with all their independent derivatives}, i.e. essentially a Taylor expansion of the fields \emph{along the lightcone} of every spacetime point $x$. Since the entire solution is thus encoded at every $x$ separately, the master fields can simply be evolved from one point to the next, via ``unfolded'' field equations. By construction, the unfolded equations are local in spacetime: they simply specify the first derivative along $x$. From this perspective, the non-locality creeps in via the twistor coordinates, which are subject to a manifestly non-local $\star$-product algebra. 

Given such complications and exotic formalism, it's tempting to sidestep the bulk HS theory, and instead work with its boundary holographic dual (which is remarkably simple -- a free vector model). However, we must keep track of HS theory's main promise (as the author sees it): to be a tractable theory of 4d quantum gravity with $\Lambda>0$. The all-important physical consequence of $\Lambda>0$ is that \emph{the boundary at infinity is unobservable}. As a result, we have no choice but to define and perform calculations \emph{within observable regions of the bulk}, such as the static patch of de Sitter space. The holographic description may still have a role to play, but as an intermediate step rather than the final answer. 

\subsection{The self-dual sector and higher-spin self-dual GR}

In both Yang-Mills and GR, the \emph{self-dual sector} is famously much simpler than the full theory, and can be used as a starting point for its construction (see e.g. the nice review \cite{Krasnov:2016emc}). In asymptotically flat spacetime ($\Lambda=0$), this self-dual sector describes MHV scattering amplitudes \cite{Bardeen:1995gk,Rosly:1996vr,Mason:2009afn}. With $\Lambda\neq 0$, the relation between the self-dual sector and asymptotic ``observables'' is much less clear, since self-duality is inconsistent with the standard boundary conditions of AdS/CFT. However, in de Sitter space ($\Lambda>0$), if we change our focus from asymptotic correlators (which, in this context, are unobservable) to scattering in the observable static patch \cite{David:2019mos,Albrychiewicz:2020ruh}, then the self-dual sector again describes the simplest, MHV-like, ``scattering amplitudes'' \cite{Albrychiewicz:2021ndv,Neiman:2023bkq}.
 
In HS gravity, the self-dual sector turns out to be an even greater simplification than in Yang-Mills and GR: while the full theory doesn't have a known formulation in standard local language, the self-dual sector does! Specifically, for $\Lambda=0$, a \emph{lightcone formulation} \cite{Ponomarev:2016lrm} of the self-dual sector, with only cubic, local vertices with helicities $(h_1,h_2,h_3)=(+s_1,+s_2,-s_3)$, was demonstrated \cite{Skvortsov:2018jea,Skvortsov:2020wtf} to be a complete, self-consistent theory (though, like all self-dual theories, it is non-unitary). The cubic lightcone vertices for $\Lambda\neq 0$ are also known \cite{Skvortsov:2018uru}, but there the expectation is that higher-order vertices will be required as well. Unfolded formulations of the self-dual sector also exist \cite{Skvortsov:2022syz,Sharapov:2022faa,Sharapov:2022awp,Didenko:2022qga}. 

The subject of the present paper is a certain \emph{restriction} of the self-dual sector of HS gravity, which is under even greater control. This restriction, known (somewhat confusingly) as HS self-dual GR, contains only \emph{nonzero even} spins, and only those cubic vertices that satisfy $h_1+h_2+h_3=2$; it can be thought of as the minimal HS extension of self-dual GR, the latter corresponding to $(h_1,h_2,h_3)=(+2,+2,-2)$. What is remarkable about this theory is that (alone among interacting HS gravities) it can be described by a covariant local action, in a straightforward extension of Krasnov's chiral formulation of GR \cite{Krasnov:2011up,Krasnov:2011pp} (see also \cite{Capovilla:1989ac,Capovilla:1990qi}). This is in contrast with the full self-dual sector, where, to obtain a local action, one must apparently sacrifice covariance in favor of the lightcone formalism. The covariant action for HS self-dual GR has been worked out \cite{Krasnov:2021nsq} for both $\Lambda=0$ and $\Lambda\neq 0$, where, in the latter case, the cubic vertices must be supplemented by quartic ones. 

Our goal will be to study HS self-dual GR and its solutions, while also exploring its connections with the constructions relevant for full HS theory and its self-dual sector: lightcone, unfolding, and the combination of spacetime with twistor-like coordinates. Our eventual aim is twofold. First, we want to build the tools for computing bulk observables in de Sitter space (such as static-patch scattering) within HS self-dual GR. Second, we want to build bridges that would allow us to extrapolate from HS self-dual GR to full HS gravity, or to its full self-dual sector.

\subsection{Summary and structure of the paper}

The results and structure of the paper are as follows. We work in the $\Lambda\neq 0$ version of HS self-dual GR. In section \ref{sec:review}, we review its definition \cite{Krasnov:2021nsq} as a non-Abelian gauge theory in 4d spacetime with an unusual ``generalized diffeomorphism'' symmetry. We observe that, upon restricting to even spins, one can interpret this theory as one of self-dual (and linearized anti-self-dual) HS fields living, with no backreaction, on a standard 4d spacetime background that solves the equations of self-dual GR. On the other hand, in section \ref{sec:6d}, we show how this 4d theory can be obtained as a gauge-fixing of a much simpler diffeomorphism-invariant gauge theory in 6 dimensions, in which the 4d spacetime coordinates $x^a$ are combined on an equal footing with 2d left-handed spinor coordinates $y^\alpha$. This construction is reminiscent of the unfolded formulation of HS theory in terms of ``master fields'' depending on $(x^a,y^\alpha,\bar y^{\dot\alpha})$, but with several important differences: the right-handed $\bar y^{\dot\alpha}$ is absent, the theory is manifestly local in both $x^a$ and $y^\alpha$, and almost no distinction exists between the roles of $x^a$ and $y^\alpha$ (aside from an orbifolding map $y^\alpha\to iy^\alpha$, which singles out ``the'' 4d spacetime $(x^a,y^\alpha)=(x^a,0)$ as its fixed submanifold). Our construction is also closely related to twistor formulations of self-dual GR in Euclidean signature, especially as presented in \cite{Herfray:2016qvg}.

Next, in section \ref{sec:lightcone_ansatz:ansatz}, we return to the 4d spacetime formulation, and gauge-fix it further to obtain a \emph{lightcone ansatz} for solutions to the covariant field equations. This generalizes the ansatz found in \cite{Neiman:2023bkq} for self-dual GR, which extended Plebanski's ``second heavenly equation'' \cite{Plebanski:1975wn} (see also \cite{Siegel:1992wd}) to $\Lambda\neq 0$. The lightcone ansatz forms a bridge between the covariant formulation \cite{Krasnov:2021nsq} of HS self-dual GR and the lightcone formulation \cite{Ponomarev:2016lrm,Skvortsov:2018jea,Skvortsov:2018uru} of the full self-dual sector of HS gravity. In section \ref{sec:lightcone_ansatz:comparing}, we point out some differences between our lightcone ansatz and Metsaev's lightcone formalism \cite{Metsaev:2005ar,Metsaev:2018xip}, which was utilized in \cite{Ponomarev:2016lrm,Skvortsov:2018jea,Skvortsov:2018uru}. In particular, we note that our ansatz allows for more general (and less symmetric) lightlike foliations of spacetime, based on the lightcones of bulk points rather than boundary ones.

In section \ref{sec:lightcone_covariance:general}, we pivot into a conceptual discussion of the lightcone formalism with $\Lambda\neq 0$, independent of the specifics of HS self-dual GR. While the $\Lambda\neq 0$ lightcone formalism is more technically involved than its $\Lambda=0$ version, we point out that this comes with an advantage: as we'll show, the $\Lambda\neq 0$ lightcone formalism is not ``trapped'' in a fixed lightlike foliation, but has a built-in mechanism for ``changing the reference frame'' and evolving along arbitrary lightlike directions in spacetime. Together with the previous remark vis. using lightcones of bulk points, this observation elevates the lightcone formalism into something akin to the unfolded language, where a master-field that encodes the fields' Taylor series along the lightcone of a spacetime point can be evolved to an arbitrary adjacent point. This analogy with unfolding is discussed in section \ref{sec:lightcone_covariance:unfolding}. In section \ref{sec:lightcone_covariance:HS}, we apply section \ref{sec:lightcone_covariance:general}'s general discussion of the lightcone formalism to the particularities of our lightcone ansatz for HS self-dual GR. Section \ref{sec:outlook} is devoted to discussion and outlook.

Throughout the paper, we use ``right-handed''/``left-handed'' as synonymous with ``self-dual''/``anti-self-dual''.

\section{Review of higher-spin self-dual GR} \label{sec:review}

In this section, we review the definition \cite{Krasnov:2021nsq} of HS self-dual GR, along with some context and notations. We begin by reviewing Krasnov's chiral formalism \cite{Krasnov:2011pp,Krasnov:2011up,Krasnov:2016emc} for self-dual and full GR.

\subsection{GR in Krasnov's formalism} \label{sec:review:GR}

We work in a curved 4d spacetime with coordinates $x^a$. Indices $(\alpha,\beta,\dots)$ and $(\dot\alpha,\dot\beta,\dots)$ are used respectively for left-handed and right-handed Weyl spinors. These are raised and lowered with the flat spinor metrics $\epsilon_{\alpha\beta}$ and $\epsilon_{\dot\alpha\dot\beta}$, according to the rules:
\begin{align}
    \zeta_\alpha = \epsilon_{\alpha\beta}\zeta^\beta \ ; \quad \zeta^\alpha = \zeta_\beta\epsilon^{\beta\alpha} \ ; \quad
    \bar\zeta_{\dot\alpha} = \epsilon_{\dot\alpha\dot\beta}\bar\zeta^{\dot\beta} \ ; \quad \bar\zeta^{\dot\alpha} = \bar\zeta_{\dot\beta}\epsilon^{\dot\beta\dot\alpha} \ .
\end{align}
With these ingredients, one can construct Cartan's formulation of GR, whose fundamental variables are 1-forms: the vielbein $e_a^{\alpha\dot\alpha}$ and the (left-handed and right-handed halves of) the spin-connection $\omega_a^{\alpha\beta} = \omega_a^{(\alpha\beta)}$ and $\omega_a^{\dot\alpha\dot\beta} = \omega_a^{(\dot\alpha\dot\beta)}$. Here, we are using spinor indices for vectors and tensors in the ``internal'' flat spacetime. Now, from e.g. the left-handed connection $\omega^{\alpha\beta}_a$ we can construct its curvature 2-form:
\begin{align}
    F^{\alpha\beta} = \frac{1}{2}F_{ab}^{\alpha\beta}dx^a\wedge dx^b \ ; \quad F_{ab}^{\alpha\beta} = 2\del_{[a}\omega_{b]}^{\alpha\beta} + \omega_{[a}^{\alpha\gamma} \omega_{b]\gamma}^\beta \ . \label{eq:F_GR}
\end{align}
From there, using the inverse vielbein $e^a_{\alpha\dot\alpha}$, we can extract the left-handed Weyl tensor (in internal spinor indices), as $\Psi^{\alpha\beta\gamma\delta} = \Psi^{(\alpha\beta\gamma\delta)} = \frac{1}{2}e^{a(\alpha|\dot\alpha}e^{b|\beta}{}_{\dot\alpha}F_{ab}^{\gamma\delta)}$. 

Now, the remarkable observation by Krasnov is that one can define GR with $\Lambda\neq 0$ using an action whose fundamental variables are the 1-form $\omega_a^{\alpha\beta}$ and the 0-form $\Psi^{\alpha\beta\gamma\delta}$, without their right-handed counterparts. The metric, the vielbein, and the right-handed spinor indices are all absent from the theory's basic formulation, and only appear as derived objects on solutions to the equations of motion. Explicitly, in form notation, the action reads:
\begin{align}
   S = \frac{i}{32\pi G}\int \left(\mathbb{1} - \tfrac{1}{2}\Psi\right)^{-1}_{\alpha\beta\gamma\delta}\,F^{\alpha\beta}\wedge F^{\gamma\delta} \ , \label{eq:S_full}
\end{align}
where $\left(\mathbb{1} - \tfrac{1}{2}\Psi\right)^{-1}$ denotes the geometric series:
\begin{align}
    \left(\left(\mathbb{1} - \tfrac{1}{2}\Psi\right)^{-1}\right)^{\alpha\beta}_{\gamma\delta} 
    = \delta^\alpha_{(\gamma} \delta^\beta_{\delta)} + \frac{1}{2}\Psi^{\alpha\beta}_{\gamma\delta} + \frac{1}{4}\Psi^{\alpha\beta}_{\zeta\xi}\Psi^{\zeta\xi}_{\gamma\delta} + \dots \ . \label{eq:geometric_series}
\end{align}
Here, we fixed the cosmological constant to $\Lambda=3$, i.e. that of de Sitter space with unit curvature radius; more generally, $\Lambda/3$ would appear as a coefficient in front of the identity in $\left(\mathbb{1} - \tfrac{1}{2}\Psi\right)^{-1}$. The statement now is that \emph{on solutions to the equations of motion} of the action \eqref{eq:S_full}, there exists a vielbein $e_a^{\alpha\dot\alpha}$ (unique up to rotations of the right-handed spinor index) that solves the vacuum Einstein equations with cosmological constant, whose corresponding left-handed spin-connection and left-handed Weyl curvature are $\omega_a^{\alpha\beta}$ and $\Psi^{\alpha\beta\gamma\delta}$ respectively.

The gauge symmetry of the action \eqref{eq:S_full} consists of diffeomorphisms and left-handed Lorentz rotations of the spinor indices $(\alpha,\beta,\dots)$. Infinitesimally, a left-handed Lorentz rotation is parameterized by $\theta^{\alpha\beta} = \theta^{(\alpha\beta)}$, and takes the form:
\begin{align}
    \delta\omega_a^{\alpha\beta} = \del_a\theta^{\alpha\beta} + \theta^{(\alpha}{}_\gamma\,\omega_a^{\beta)\gamma} \ ; \quad 
    \delta\Psi^{\alpha\beta\gamma\delta} = 2\theta^{(\alpha}{}_\zeta\Psi^{\beta\gamma\delta)\zeta} \ . \label{eq:GR_gauge}
\end{align}
An infinitesimal diffeomorphism is parameterized by a vector $\xi^a$, and takes the form:
\begin{align}
    \delta\omega_a^{\alpha\beta} = \xi^b F_{ba}^{\alpha\beta} \ ; \quad 
    \delta\Psi^{\alpha\beta\gamma\delta} = \xi^a\!\left(\del_a\Psi^{\alpha\beta\gamma\delta} - 2\omega_a{}^{(\alpha}{}_\zeta\Psi^{\beta\gamma\delta)\zeta} \right) \ . \label{eq:GR_diffeo}
\end{align}
Here, we made the diffeomorphism covariant under the left-handed Lorentz symmetry \eqref{eq:GR_gauge} by combining it with a rotation $\theta^{\alpha\beta} = -\xi^a\omega_a^{\alpha\beta}$.

\subsection{Self-dual GR} \label{sec:review:SD_GR}

The chiral formulation \eqref{eq:S_full} of full GR contains \emph{self-dual GR} as a simple subsector. Self-dual GR is characterized by the Einstein equations \emph{together with} the vanishing of the left-handed Weyl curvature $\Psi^{\alpha\beta\gamma\delta}$. To descend into this sector, we pick out the lowest non-trivial power of $\Psi^{\alpha\beta\gamma\delta}$ in the geometric series \eqref{eq:geometric_series}. The zeroth power is trivial, leading to a topological action $S\sim \int F_{\alpha\beta}\wedge F^{\alpha\beta}$. Self-dual GR is in fact obtained by picking out the \emph{first} power:
\begin{align}
    S = \frac{i}{64\pi G}\int \Psi_{\alpha\beta\gamma\delta}\,F^{\alpha\beta}\wedge F^{\gamma\delta} \ . \label{eq:S_SD_GR}
\end{align}
Varying this action with respect to $\Psi^{\alpha\beta\gamma\delta}$, we obtain the non-linear field equation:
\begin{align}
    F^{(\alpha\beta}\wedge F^{\gamma\delta)} = 0 \ , \label{eq:SD_GR_equation}
\end{align}
in which the only fundamental variable is $\omega_a^{\alpha\beta}$. On solutions of this equation, there again exists a vielbein $e_a^{\alpha\dot\alpha}$ compatible with $\omega_a^{\alpha\beta}$, which solves the Einstein equation and has vanishing left-handed Weyl curvature. In fact, the two are related simply via:
\begin{align}
    F_{ab}^{\alpha\beta} = 2e_{[a}^{\alpha\dot\alpha} e^{\beta}_{b]\dot\alpha} \ .
\end{align}
The variable $\Psi_{\alpha\beta\gamma\delta}$ in \eqref{eq:S_SD_GR} is no longer the left-handed Weyl curvature of $e_a^{\alpha\dot\alpha}$ and $\omega_a^{\alpha\beta}$ (which vanishes). Instead, it represents a \emph{linearized anti-self-dual perturbation} on the non-linear self-dual background defined by $\omega_a^{\alpha\beta}$. The field equation for its linearized propagation is obtained by varying \eqref{eq:S_SD_GR} with respect to $\omega_a^{\alpha\beta}$. The gauge symmetry of the action \eqref{eq:S_SD_GR} is once again given by the left-handed Lorentz rotations \eqref{eq:GR_gauge} and diffeomorphisms \eqref{eq:GR_diffeo}.

\subsection{Higher-spin self-dual GR} \label{sec:review:HS_SD_GR}

Let us now introduce the main topic of this paper -- the higher-spin generalization \cite{Krasnov:2021nsq} of the self-dual GR action \eqref{eq:S_SD_GR}. We define HS versions of the fundamental variables $\omega_a^{\alpha\beta},\Psi^{\alpha\beta\gamma\delta}$ and gauge parameters $\theta^{\alpha\beta},\xi^a$ as:
\begin{align}
    \omega_a(x,y) &= \sum_s \frac{y_{\alpha_1}\dots y_{\alpha_{2s-2}}}{(2s-2)!}\,\omega_a^{\alpha_1\dots\alpha_{2s-2}}(x) \ ; \quad
    \Psi(x,y) = \sum_s \frac{y^{\alpha_1}\dots y^{\alpha_{2s}}}{(2s)!}\,\Psi_{\alpha_1\dots\alpha_{2s}}(x) \ ; \label{eq:fields_xy} \\
    \theta(x,y) &= \sum_s \frac{y_{\alpha_1}\dots y_{\alpha_{2s-2}}}{(2s-2)!}\,\theta^{\alpha_1\dots\alpha_{2s-2}}(x) \ ; \quad
    \xi^a(x,y) = \sum_s \frac{y^{\alpha_1}\dots y^{\alpha_{2s-4}}}{(2s-4)!}\,\xi^a_{\alpha_1\dots\alpha_{2s-4}}(x) \ . \label{eq:gauge_parameters_xy}
\end{align}
Here, alongside the dependence on spacetime coordinates $x^a$, our fields $\omega_a,\Psi$ and gauge parameters $\theta,\xi^a$ depend on a left-handed spinor $y^\alpha$, which compactly packages an infinite tower of spins. The spin $s$ in \eqref{eq:fields_xy}-\eqref{eq:gauge_parameters_xy} takes positive even values $s=2,4,6,\dots$, where $s=2$ corresponds to the GR fields and gauge parameters from sections \ref{sec:review:GR}-\ref{sec:review:SD_GR}. In this, we depart from \cite{Krasnov:2021nsq}, where odd spins were included as well.

We define the curvature 2-form $F$ of the connection $\omega_a$ as:
\begin{align}
    F = \frac{1}{2}F_{ab}(x,y)dx^a\wedge dx^b \ ; \quad F_{ab} = 2\del_{[a}\omega_{b]} - \frac{1}{2}\del_\gamma\omega_{[a} \del^\gamma\omega_{b]} \ , \label{eq:F}
\end{align}
where $\del_\alpha \equiv \del/\del y^\alpha$ denotes differentiation with respect to the spinor coordinates, and $\del^\alpha\equiv \epsilon^{\beta\alpha}\del_\beta$ is its raised-index version. The curvature \eqref{eq:F} can again be decomposed into spins as:
\begin{align}
    F_{ab}(x,y) = \sum_s \frac{y_{\gamma_1}\dots y_{\gamma_{2s-2}}}{(2s-2)!}\,F_{ab}^{\gamma_1\dots\gamma_{2s-2}}(x) \ ,
\end{align}
with $s=2$ corresponding to the GR curvature \eqref{eq:F_GR}. 

The action of HS self-dual GR is now given by:
\begin{align}
    S = \frac{i}{16\pi G}\int \left.\Psi\!\left(x,\frac{\del}{\del y}\right) F(x,y)\wedge F(x,y) \right|_{y^\alpha=0} \ , \label{eq:S_HS_SD_GR}
\end{align}
which reduces to \eqref{eq:S_SD_GR} if we restrict all fields to $s=2$. Varying the action w.r.t. $\Psi$ again gives a non-linear field equation for $\omega_a$:
\begin{align}
    F\wedge F = 0 \ , \label{eq:F_wedge_F}
\end{align}
while varying w.r.t. $\omega_a$ gives a linear equation for the propagation of $\Psi$ over the non-linear background defined by $\omega_a$. 

The action \eqref{eq:S_HS_SD_GR} is invariant under internal gauge transformations parameterized by $\theta(x,y)$:
\begin{align}
    \delta\omega_a(x,y) = \del_a\theta + \frac{1}{2}\,\del_\beta\theta\,\del^\beta\omega_a \ ; \quad \delta\Psi(x,y) = \frac{y^\alpha}{2}\,(\del_\alpha\theta)\!\left(x,\frac{\del}{\del y}\right)\Psi(x,y) \equiv \theta\circ\Psi \ ,
    \label{eq:internal_gauge}
\end{align}
as well as under HS diffeomorphisms parameterized by $\xi^a(x,y)$:
\begin{align}
    \delta\omega_a(x,y) = \xi^b F_{ba} \ ; \quad \delta\Psi(x,y) = \xi^a\!\left(x,\frac{\del}{\del y}\right)(\del_a\Psi - \omega_a\circ\Psi) \ . \label{eq:diffeo}
\end{align}
In words, the definition of $\theta\circ\Psi$ in \eqref{eq:internal_gauge} is to take the derivative $\del_\alpha\theta(x,y)$, then substitute $y^\beta\to\del^\beta$, then act with the resulting operator on $\Psi(x,y)$, and finally multiply by $y^\alpha/2$.

\subsection{Two observations} \label{sec:review:observations}

We now make two simple observations concerning the theory \eqref{eq:S_HS_SD_GR}, which are absent from the original paper \cite{Krasnov:2021nsq}. 

\subsubsection{No quintic vertex}

Our first observation concerns the degree of the interaction vertices in \eqref{eq:S_HS_SD_GR}. Naively, we have up to quintic vertices, since each factor of $F$ in \eqref{eq:S_HS_SD_GR} is quadratic in the fundamental field $\omega_a$, via \eqref{eq:F}. However, the quintic vertex, or, equivalently, the quartic term in the field equation \eqref{eq:F_wedge_F}, vanishes identically. The reason is that these terms contain a wedge product $\del^\alpha\omega_{[a}\del^\beta\omega_b\del^\gamma\omega_{c]}$, which vanishes because the spinor space is only 2-dimensional (and therefore $\del^\alpha\omega_a$ denotes only \emph{two} linearly independent 1-forms on the spacetime). Explicitly, we can expand $F\wedge F$ as:
\begin{align}
    F\wedge F = \left(\del_a\omega_b\,\del_c\omega_d + \frac{1}{2}\del_a\omega_b\del^\alpha\omega_c\,\del_\alpha\omega_d\right)dx^a\wedge dx^b\wedge dx^c\wedge dx^d \ . \label{eq:F_wedge_F_explicit}
\end{align}
Thus, the action \eqref{eq:S_HS_SD_GR} has only cubic and quartic terms.

\subsubsection{No backreaction on self-dual GR spacetime} \label{sec:review:observations:no_backreaction}

Our second observation is that the theory describes a standard self-dual GR spacetime, encoded by $\omega_a^{\alpha\beta}$ with curvature \eqref{eq:F_GR} subject to the field equation \eqref{eq:SD_GR_equation}, on which all other fields (i.e. the connections $\omega_a^{\alpha_1\dots\alpha_{2s-2}}$ for $s>2$ and the left-handed field strengths $\Psi^{\alpha_1\dots\alpha_{2s}}$ for all $s$) propagate and interact with no backreaction. Moreover, the self-dual GR spacetime described by $\omega_a^{\alpha\beta}$ is unaffected by the $s>2$ components of the gauge transformations $\theta^{\alpha_1\dots\alpha_{2s-2}}$ and $\xi^a_{\alpha_1\dots\alpha_{2s-4}}$. In more detail, our observation consists of the following points:
\begin{enumerate}
    \item In \eqref{eq:F}, the $s=2$ curvature $F_{ab}^{\alpha\beta}$ depends solely on the $s=2$ connection $\omega_a^{\alpha\beta}$.
    \item When expanding the field equation \eqref{eq:F_wedge_F} in powers of $y^\alpha$, the lowest non-vanishing power is 4, and its coefficient is just the self-dual GR field equation \eqref{eq:SD_GR_equation}.
    \item In \eqref{eq:internal_gauge}-\eqref{eq:diffeo}, the components of $\theta,\xi^a$ with a given spin only affect the components of $\omega_a$ with the \emph{same spin or higher}. As a corollary, the $s>2$ gauge transformations do not affect the $s=2$ connection $\omega_a^{\alpha\beta}$.
\end{enumerate}
Note that points 2 and 3 fail to hold if we allow odd spins.

\section{Diffeomorphism-invariant formulation in 6d} \label{sec:6d}

In this section, we present a new formulation of HS self-dual GR, which is diffeomorphism-invariant not just in 4d spacetime, but in the 6d space coordinatized by $X^I\equiv (x^a,y^\alpha)$. There are two main hints that such a formulation should exist:
\begin{itemize}
    \item Already in the original formulation \cite{Krasnov:2021nsq}, the curvature \eqref{eq:F} and field equation \eqref{eq:F_wedge_F} are manifestly local not only in spacetime, but also in the space of spinors $y^\alpha$.
    \item The $y$-dependence of the generalized diffeomorphisms $\xi^a(x,y)$ looks as if they are ``trying'' to be embedded into full 6d diffeomorphisms $\Xi^I(X)$.
\end{itemize}
In fact, such a 6d formulation was already attempted \cite{Herfray:2022prf}, but in the end didn't produce an interpretation of $\xi^a(x,y)$ in terms of 6d diffeomorphisms. Our construction will be somewhat different.

\subsection{Construction} \label{sec:6d:construction}

We begin by Fourier-transforming $\Psi(x,y)$ with respect to $y^\alpha$:
\begin{align}
    \tilde\Psi(x,y) \equiv \int\frac{d^2u}{(2\pi)^2}\,e^{iu_\alpha y^\alpha}\Psi(x,u) \ , \label{eq:Psi_Fourier}
\end{align},
where the measure $d^2u$ is defined as: 
\begin{align}
    d^2u\equiv \frac{1}{2}\epsilon_{\alpha\beta}\,du^\alpha\wedge du^\beta = \frac{1}{2}du^\alpha\wedge du_\alpha \ . \label{eq:measure}
\end{align}
This converts the non-local $y^\alpha$-space structure of the action \eqref{eq:S_HS_SD_GR} into a local integral:
\begin{align}
    S = \frac{i}{16\pi G}\int \tilde\Psi(x,y)\,F(x,y)\wedge F(x,y)\wedge d^2y \ , \label{eq:S_almost_6d}
\end{align}
where the integration is now over the 6d space $(x^a,y^\alpha)$. Our next step towards 6d covariance is to extend the 4-dimensional 1-form $\omega_a$ into a 6-dimensional 1-form:
\begin{align}
    \Omega \equiv \Omega_I dX^I \equiv \omega_a(x,y) dx^a + y_\alpha dy^\alpha \ , \label{eq:Omega}
\end{align}
whose exterior derivative reads:
\begin{align}
    d\Omega = \del_a\omega_b\,dx^a\wedge dx^b + \del_\alpha\omega_a\,dy^\alpha\wedge dx^a - 2d^2y \ , \label{eq:d_Omega}
\end{align}
with $d^2y$ defined as in \eqref{eq:measure}. From this, we can construct a 6d top form $d\Omega\wedge d\Omega\wedge d\Omega$. Computing it from \eqref{eq:d_Omega}, we get two kinds of contributions: one with two $dx^a\wedge dx^b$ factors and one $d^2y$ factor, and another with one $dx^a\wedge dx^b$ factor and two $dy^\alpha\wedge dx^a$ factors. Altogether, we get simply:
\begin{align}
    d\Omega\wedge d\Omega\wedge d\Omega = -6F\wedge F\wedge d^2y \ ,
\end{align}
where we used the cubic expression \eqref{eq:F_wedge_F_explicit} for $F\wedge F$. The action \eqref{eq:S_almost_6d} thus becomes:
\begin{align}
    S = -\frac{i}{96\pi G}\int \tilde\Psi\,d\Omega\wedge d\Omega\wedge d\Omega \ . \label{eq:S_6d}
\end{align}
This is manifestly invariant under internal gauge transformations parameterized by $\theta(X)$ and 6d diffeomorphisms parameterized by $\Xi^I(X)\equiv (\xi^a,\xi^\alpha)$:
\begin{align}
    \delta\Omega_I = \del_I\theta + 2\,\Xi^J\del_{[J}\Omega_{I]} \ ; \quad \delta\Psi = \Xi^I\del_I\Psi \ . \label{eq:6d_gauge}
\end{align}
With respect to these gauge transformations, our ansatz \eqref{eq:Omega} for $\Omega$, i.e. the constraint that its $dy^\alpha$ components are given by $y_\alpha$, can be viewed as a \emph{partial gauge fixing}. Indeed, for a generic $\Omega$, one can first use a diffeomorphism to fix the $d^2y$ component of $d\Omega$ to $-2$ as in \eqref{eq:d_Omega}, and then use an internal gauge transformation to fix the $dy^\alpha$ components of $\Omega$ completely.

Conversely, we can ask which 6d gauge transformations \eqref{eq:6d_gauge} preserve the ansatz \eqref{eq:Omega}. This can be found by setting the $dy^\alpha$ components of the variation $\delta\Omega$ to zero, which fixes $\xi^\alpha$ in terms of $\xi^a$ and $\theta$:
\begin{align}
    \xi^\alpha = \frac{1}{2}(\xi^b\del^\alpha\omega_b - \del^\alpha\theta) \ . \label{eq:xi_alpha}
\end{align}
Under this condition, the 6d gauge transformations \eqref{eq:6d_gauge} take the form:
\begin{align}
    \delta\omega_a &= \del_a\theta + \frac{1}{2}\,\del_\beta\theta\,\del^\beta\omega_a + \xi^b F_{ba} \ ; \label{eq:delta_omega_6d} \\
    \delta\tilde\Psi &= \frac{1}{2}\,\del_\alpha\theta\,\del^\alpha\tilde\Psi + \xi^a\left(\del_a\tilde\Psi - \frac{1}{2}\,\del_\beta\omega_a\,\del^\beta\tilde\Psi \right) \ . \label{eq:delta_Psi_6d}
\end{align}
This reproduces the gauge transformations \eqref{eq:internal_gauge}-\eqref{eq:diffeo} in the original formulation, with the transformations $\delta\Psi$ from \eqref{eq:internal_gauge}-\eqref{eq:diffeo} replaced by the simpler (and $y$-local) eq. \eqref{eq:delta_Psi_6d} thanks to the Fourier transform \eqref{eq:Psi_Fourier}. Thus, the original gauge symmetries \eqref{eq:internal_gauge}-\eqref{eq:diffeo} are precisely the subset of the 6d gauge symmetries \eqref{eq:6d_gauge} that preserves the ansatz \eqref{eq:Omega}. 

The final ingredient in our construction is to make explicit the restriction to nonzero even spins in \eqref{eq:fields_xy}-\eqref{eq:gauge_parameters_xy}. This can be implemented in the 6d picture by imposing an orbifolding map:
\begin{align}
    (x^a,y^\alpha)\to (x^a,iy^\alpha) \quad \Longrightarrow \quad \Omega\to -\Omega \, ; \ \tilde\Psi\to -\tilde\Psi \, ; \ \theta\to -\theta \, ; \ (\xi^a,\xi^\alpha)\to (\xi^a,i\xi^\alpha) \ . \label{eq:orbifold}
\end{align}
This reproduces the allowed spins in \eqref{eq:fields_xy}-\eqref{eq:gauge_parameters_xy}, with one exception: it fails to rule out a spin-0 component in $\Psi$. However, this is harmless: such a component will not contribute to the action \eqref{eq:S_HS_SD_GR}, since $F$ vanishes at $y^\alpha=0$ (note that this last point relies on the restriction to even spins). 

The orbifolding \eqref{eq:orbifold} is essential to section \ref{sec:review:observations:no_backreaction}'s observation that the HS theory ``lives on'' a standard 4d self-dual GR spacetime. Indeed, the 4d spacetime itself is the singular manifold $(x^a,y^\alpha)=(x^a,0)$ of the map \eqref{eq:orbifold}. On this 4d manifold, $\Omega$ itself vanishes; the 4d geometry $\omega_a^{\alpha\beta}(x)$ is then encoded in the leading non-trivial derivatives of (the 4d pullback of) $\Omega$ along the transverse directions $y^\alpha$.

\subsection{Comparison to the Euclidean twistor picture of self-dual GR}

The 6d construction presented above is closely related to the famous ``non-linear graviton'' formulation of self-dual GR in twistor space \cite{Penrose:1976js,Ward:1980am,Herfray:2016qvg}. Indeed, in Euclidean signature, twistor space can be identified with the left-handed spinor bundle on spacetime, i.e. precisely the space $(x^a,y^\alpha)$. Moreover, if we restrict $\omega_a(x,y)$ to only its spin-2 component $\frac{1}{2}y_\alpha y_\beta\omega_a^{\alpha\beta}(x)$, then our 1-form $\Omega$ from \eqref{eq:Omega} becomes precisely the 1-form denoted as $\tau$ in the twistor literature. As usual, in theories with fixed spin it makes sense to work in \emph{projective} twistor space, i.e. to quotient out the rescalings $y^\alpha\to \rho y^\alpha$, whereas in HS theory it makes sense to stay in the non-projective space, with different weights under $y^\alpha\to \rho y^\alpha$ encoding different spins. 

Perhaps not surprisingly, as we saw in eq. \eqref{eq:6d_gauge}, our HS construction in section \ref{sec:6d:construction} has a higher degree of gauge symmetry than the spin-2 ``non-linear graviton'': 
\begin{enumerate}
    \item It admits arbitrary 6d diffeomorphisms, which treat $x^a$ and $y^\alpha$ on an equal footing (subject to the orbifolding map \eqref{eq:orbifold}).
    \item It has an internal Abelian gauge symmetry $\delta\Omega = d\theta$, which elegantly replaces the non-Abelian internal gauge symmetry \eqref{eq:internal_gauge} of the 4d formulation. In particular, this symmetry is manifest in the 6-form $d\Omega\wedge d\Omega\wedge d\Omega$ that appears in our action \eqref{eq:S_6d}, as opposed to the 5-form $\tau\wedge d\tau\wedge d\tau$ that appears in the spin-2 theory.
\end{enumerate}
On the flipside, this higher degree of symmetry means that our construction lacks some familiar structures from the spin-2 story:
\begin{enumerate}
    \item The freedom of 6d diffeomorphisms means that our space $(x^a,y^\alpha)$ is no longer a \emph{bundle} of spinors $y^\alpha$ over spacetime $x^a$. In other words, the HS symmetries do not respect the bundle projection $(x^a,y^\alpha)\to x^a$, which gets replaced by the (much weaker) orbifolding map \eqref{eq:orbifold}. 
    \item This lack of separation between $x^a$ and $y^\alpha$ prevents us from treating $x^a$ as real and $y^\alpha$ as complex, as should be the case in Euclidean signature. In this paper, we'll simply be content with treating both $x^a$ and $y^\alpha$ as complex; this makes sense from the perspective of Lorentzian signature, where any self-dual theory is always complexified. One could also consider both $x^a$ and $y^\alpha$ real, as in $(2,2)$ signature; however, we then lose the elegant encoding \eqref{eq:orbifold} of the restriction to even spins.
\end{enumerate}
The loss of these bundle and reality structures means that the space $(x^a,y^\alpha)$ \emph{can no longer be identified} with Euclidean-signature twistor space. This accounts for the differences between our 6d construction and the one in \cite{Herfray:2022prf}, which insisted on maintaining this identification and the requisite structures.

\section{Lightcone ansatz for HS self-dual GR} \label{sec:lightcone_ansatz}

In this section, we present our lightcone ansatz for HS self-dual GR. This generalizes our previous spin-2 result \cite{Neiman:2023bkq}, which in turn was an extension of Plebanski's ``second heavenly equation'' \cite{Plebanski:1975wn} to $\Lambda\neq 0$. Here and for the remainder of the paper, we revert to the original 4d definition \cite{Krasnov:2021nsq} of HS self-dual GR, as introduced in section \ref{sec:review:HS_SD_GR}.

\subsection{Ansatz and field equations} \label{sec:lightcone_ansatz:ansatz}

Our goal is to write a lightcone ansatz for the right-handed solutions $\omega_a(x,y)$ to the field equation \eqref{eq:F_wedge_F}. Such an ansatz reduces the covariant field $\omega_a(x,y)$ to a single (scalar) physical degree of freedom for each spin, subject to appropriate scalar field equations. We can then plug the ansatz back into the action \eqref{eq:S_HS_SD_GR}, and obtain similar scalar field equations for the linearized left-handed degrees of freedom encoded in $\Psi(x,y)$.

We will use Poincare coordinates $x^a=(t,\mathbf{x})$, as well as the Pauli matrices $\sigma_a^{\alpha\dot\alpha},\sigma^a_{\alpha\dot\alpha}$, which satisfy:
\begin{align}
    \sigma_a^{\alpha\dot\alpha} \sigma_{b\alpha\dot\alpha} = -2\eta_{ab} \ ; \quad \sigma_a^{\alpha\dot\alpha}\sigma^b_{\alpha\dot\alpha} = -2\delta_a^b \ ; \quad
    \sigma^a_{\alpha\dot\alpha}\sigma^{b\alpha\dot\alpha} = -2\eta^{ab} \ ,
\end{align}
with $\eta_{ab}$ the Minkowski metric $\eta_{ab}dx^adx^b = -dt^2+\mathbf{dx}^2$, and $\eta^{ab}$ its inverse. 

Our lightcone ansatz has two more ingredients: an arbitrary fixed spinor $q^\alpha$ that defines the preferred lightlike direction, and a generating function $\phi(x^a,u)$ with scalar variable $u$ that encodes the dynamical right-handed degree of freedom for each spin:
\begin{align}
    \phi(x,u) = \sum_s u^{s-1} \phi^{(s)}(x) \ . \label{eq:phi_decomposition}
\end{align}
In terms of these ingredients, our ansatz for the connection $\omega_a$ reads:
\begin{align}
 \omega_a(x,y) &= -\frac{1}{2}\sigma_a^{\alpha\dot\alpha}\omega_{\alpha\dot\alpha}(x,y) \ ; \label{eq:ansatz_1} \\
 \omega_{\alpha\dot\alpha}(x,y) &= -y_\alpha y^\beta\del_{\beta\dot\alpha}\ln t + q_\alpha q^\beta\del_{\beta\dot\alpha}\phi(x,u)\big|_{u\,=\,\langle qy\rangle^2/t} \ , \label{eq:ansatz_2}
\end{align}
where we defined the shorthands $\del_{\alpha\dot\alpha} \equiv \sigma^a_{\alpha\dot\alpha}\del_a$ and $\langle qy\rangle \equiv q_\alpha y^\alpha$, and we substitute $u=\langle qy\rangle^2/t$ \emph{after} taking the spacetime gradient $\del_{\beta\dot\alpha}$. Note that for later convenience, there is a factor of 2 between the spin-2 scalar $\phi^{(2)}$ as defined in \eqref{eq:phi_decomposition}-\eqref{eq:ansatz_2} and the corresponding definition in \cite{Neiman:2023bkq}.

The first term in \eqref{eq:ansatz_2} is the (purely spin-2) left-handed spin-connection of pure de Sitter space, with metric $\eta_{ab}/t^2$ \cite{Krasnov:2011up}. The second term then describes the deviation (of all spins $s\geq 2$) from this pure de Sitter solution. We note the elegant algebraic similarity between the two terms. The curvature \eqref{eq:F} of the connection \eqref{eq:ansatz_1}-\eqref{eq:ansatz_2} reads:
\begin{align}
  \begin{split}
     F_{ab}(x,y) &= \frac{1}{4}\sigma_a^{\alpha\dot\alpha}\sigma_b^{\beta\dot\beta}\big(\epsilon_{\dot\alpha\dot\beta}F_{\alpha\beta}(x,y) + \epsilon_{\alpha\beta}F_{\dot\alpha\dot\beta}(x,y) \big) \ ; \\
     F_{\alpha\beta}(x,y) &= \frac{2y_\alpha y_\beta}{t^2} - q_\alpha q_\beta\Box\phi 
       + \frac{2q_{(\alpha} y_{\beta)} q^\gamma q^\delta}{\langle qy\rangle}\,\del_\gamma{}^{\dot\beta}\ln t\,\del_{\delta\dot\beta}(u\del_u\phi) \ ; \\
     F_{\dot\alpha\dot\beta}(x,y) &= q_\alpha q_\beta\big( {-\del^\alpha{}_{\dot\alpha}\del^\beta{}_{\dot\beta}\phi} + 2\del^\alpha{}_{(\dot\alpha}\ln t\,\del^\beta{}_{\dot\beta)}(u\del_u\phi) \big) \ . 
  \end{split} \label{eq:F_ansatz}
\end{align}
Here, $\Box$ is the flat d'Alembertian $\Box\equiv \eta^{ab}\del_a\del_b$; the operator $u\del_u\equiv u(\del/\del u)$ acts on each spin-$s$ component of $\phi$ as multiplication by $s-1$; and, as before, we substitute $u=\langle qy\rangle^2/t$ \emph{after} taking all the derivatives. Note that, while the curvature formula \eqref{eq:F} is quadratic in the potential $\omega_a$, the result \eqref{eq:F_ansatz} is \emph{linear} in the deformation $\phi$. This happens because the $\phi$ term in \eqref{eq:ansatz_2} depends on $y^\alpha$ only through the product $\langle qy\rangle$; as a result, the would-be quadratic contribution to $\del_\alpha\omega\,\del^\alpha\omega$ takes the form $\sim q_\alpha q^\alpha$, which vanishes.

Let us now plug \eqref{eq:F_ansatz} into the LHS of the field equation \eqref{eq:F_wedge_F}. In Poincare coordinates, we can write the 4-form $F\wedge F$ as: 
\begin{align}
    F\wedge F = \frac{i}{4}(F_{\dot\alpha\dot\beta}F^{\dot\alpha\dot\beta} - F_{\alpha\beta}F^{\alpha\beta})d^4x \ ; \quad d^4x \equiv \frac{1}{24}\epsilon_{abcd}dx^a\wedge dx^b\wedge dx^c\wedge dx^d \ ,
    \label{eq:F_F_Poincare}
\end{align}
where $F_{\dot\alpha\dot\beta}F^{\dot\alpha\dot\beta} - F_{\alpha\beta}F^{\alpha\beta}$ evaluates in our ansatz as:
\begin{align}
  F_{\dot\alpha\dot\beta}F^{\dot\alpha\dot\beta} - F_{\alpha\beta}F^{\alpha\beta} = \frac{4}{t}\,\Box(u\phi) 
    + q^\alpha q^\beta q^\gamma q^\delta\,\del_{\alpha\dot\alpha}\del_{\beta\dot\beta}\phi \big( \del_\gamma{}^{\dot\alpha}\del_\delta{}^{\dot\beta}\phi 
    - 4\,\del_\gamma{}^{\dot\alpha}\ln t\,\del_\delta{}^{\dot\beta}(u\del_u\phi) \big) \ . \label{eq:F_F_ansatz}
\end{align}
The field equation \eqref{eq:F_wedge_F} thus becomes:
\begin{align}
    \Box(u\phi) = q^\alpha q^\beta q^\gamma q^\delta\,\del_{\alpha\dot\alpha}\del_{\beta\dot\beta}\phi\left(-\frac{t}{4}\,\del_\gamma{}^{\dot\alpha}\del_\delta{}^{\dot\beta}\phi 
      + \del_\gamma{}^{\dot\alpha}t\,\del_\delta{}^{\dot\beta}(u\del_u\phi) \right) \ . \label{eq:phi_equation}
\end{align}
Crucially, all the $y^\alpha$-dependence is packaged as dependence on the scalar $u = \langle qy\rangle^2/t$. We thus have one \emph{scalar} field equation for each spin, as desired. Explicitly, we can obtain field equations for the scalar fields $\phi^{(s)}(x)$ of different spins by expanding \eqref{eq:phi_equation} in powers of $u$:
\begin{align}
    \Box\phi^{(s)} = q^\alpha q^\beta q^\gamma q^\delta \!\! \sum_{s_1+s_2 = s+2} \!\! \del_{\alpha\dot\alpha}\del_{\beta\dot\beta}\phi^{(s_1)}
      \left(-\frac{t}{4}\,\del_\gamma{}^{\dot\alpha}\del_\delta{}^{\dot\beta}\phi^{(s_2)} + (s_2-1)\del_\gamma{}^{\dot\alpha}t\,\del_\delta{}^{\dot\beta}\phi^{(s_2)} \right) \ . \label{eq:phi_s_equation}
\end{align}
For $s=2$, this reproduces the lightcone field equation for self-dual GR found in \cite{Neiman:2023bkq} (up to the overall factor of 2 mentioned above). Finally, we can plug \eqref{eq:F_F_Poincare}-\eqref{eq:F_F_ansatz} into the covariant action \eqref{eq:S_HS_SD_GR}, to obtain a lightcone action:
\begin{align}
  \begin{split}
    S ={}& \frac{1}{16\pi G}\int d^4x \sum_s \psi^{(s)}\left(\Box\phi^{(s)} \vphantom{\sum_{s_1+s_2 = s+2}} \right. \\
      &\left.{} + q^\alpha q^\beta q^\gamma q^\delta \!\! \sum_{s_1+s_2 = s+2} \del_{\alpha\dot\alpha}\del_{\beta\dot\beta}\phi^{(s_1)}
        \left(\frac{t}{4}\,\del_\gamma{}^{\dot\alpha}\del_\delta{}^{\dot\beta}\phi^{(s_2)} - (s_2-1)\del_\gamma{}^{\dot\alpha}t\,\del_\delta{}^{\dot\beta}\phi^{(s_2)} \right) \right) \ ,
  \end{split} \label{eq:S_lightcone}
\end{align}
where the non-linear right-handed degrees of freedom $\phi^{(s)}(x)$ are now joined by linearized left-handed ones $\psi^{(s)}(x)$:
\begin{align}
    \psi^{(s)}(x) = -\frac{q^{\alpha_1}\dots q^{\alpha_{2s}}}{t^{s+1}}\,\Psi_{\alpha_1\dots\alpha_{2s}}(x) \ . \label{eq:psi}
\end{align}
Just like in the spin-2 case \cite{Neiman:2023bkq}, the lightcone action \eqref{eq:S_lightcone} has only cubic interaction vertices. Its variation w.r.t. $\psi^{(s)}$ yields the quadratic field equation \eqref{eq:phi_s_equation} for $\Box\phi^{(s)}$. Similarly, the variation w.r.t. $\phi^{(s)}$ yields a field equation for $\Box\psi^{(s)}$, whose RHS is bilinear in $\phi^{(s_1)}$ and $\psi^{(s_2)}$. The main difference is that the RHS for $\Box\phi^{(s)}$ gets contributions from $s_1+s_2=s+2$, i.e. from a finite collection of spins $s_1,s_2\leq s$, whereas the RHS for $\Box\psi^{(s)}$ gets contributions from $s_2-s_1=s-2$, i.e. from an infinite tower of spins.

\subsection{Comparing to Metsaev's lightcone formalism} \label{sec:lightcone_ansatz:comparing}

The cubic lightcone action \eqref{eq:S_lightcone} is very similar in spirit and structure to the general ones described by Metsaev in \cite{Metsaev:2018xip}, and applied to HS gravity in \cite{Skvortsov:2018uru}. However, there are several differences, which we pause to discuss here. Our comments will apply equally to self-dual GR as analyzed in \cite{Neiman:2023bkq}, and to HS self-dual GR as analyzed in section \ref{sec:lightcone_ansatz:ansatz} above.

\subsubsection{Cubic-exactness}

The first point to emphasize is that our cubic action \eqref{eq:S_lightcone} is exact, i.e. it doesn't require any higher-order interactions for its consistency. This should be contrasted with both the covariant action \eqref{eq:S_HS_SD_GR}, which explicitly contains quartic vertices, and also with the general analysis of \cite{Metsaev:2018xip}, where only cubic-order consistency was considered, with the expectation that consistency at higher orders will require higher-order vertices. Note that for $\Lambda=0$, the situation is simpler: there, both the covariant action for HS self-dual GR \cite{Krasnov:2021nsq} and the exact lightcone action for self-dual HS gravity \cite{Ponomarev:2016lrm,Skvortsov:2018jea,Skvortsov:2020wtf} require only cubic vertices.

\subsubsection{Lightcones of bulk points are allowed} \label{sec:lightcone_ansatz:comparing:bulk}

Another observation is that our lightcone formalism from section \ref{sec:lightcone_ansatz:ansatz} is, in a sense, more general than that of \cite{Metsaev:2018xip,Skvortsov:2018uru}. Specifically, the Poincare/lightcone coordinates in \cite{Metsaev:2018xip,Skvortsov:2018uru} are constructed so that the preferred lightlike vector $q^\alpha\bar q^{\dot\alpha}$ ($\del_-$ in the notations of \cite{Metsaev:2018xip,Skvortsov:2018uru}) is orthogonal to the gradient $\del_{\alpha\dot\alpha}t$ of the metric's scale factor ($\del_z$ in the notations of \cite{Metsaev:2018xip,Skvortsov:2018uru}). Geometrically, this means that the Poincare-coordinate ``null hyperplanes'' $q_\alpha\bar q_{\dot\alpha}x^{\alpha\dot\alpha} = \const$ are the lightcones of points on the spacetime's conformal \emph{boundary} (just like actual null hyperplanes in Minkowski space). In contrast, in our section \ref{sec:lightcone_ansatz:ansatz}, we never made the assumption $q^\alpha\bar q^{\dot\alpha}\del_{\alpha\dot\alpha}t = 0$; in fact, we've been working with $\Lambda>0$, where $\del_{\alpha\dot\alpha}t$ is timelike, and so is not orthogonal to \emph{any} lightlike vector (unless the vector is complex). The geometric significance of $q^\alpha\bar q^{\dot\alpha}\del_{\alpha\dot\alpha}t\neq 0$ is that the ``null hyperplanes'' $q_\alpha\bar q_{\dot\alpha}x^{\alpha\dot\alpha} = \const$ are now the lightcones not of boundary points, but of \emph{bulk} points (specifically, of points on the cosmological horizon that forms the ``lightlike infinity'' of the Poincare coordinates). 

In the general analysis of \cite{Metsaev:2018xip}, this more general option was avoided for good reason: the lightcones of boundary points break fewer of the (A)dS symmetries, allowing more of the symmetry generators to be treated as ``kinematical''. Nevertheless, we see that the \emph{specific} lightcone action \eqref{eq:S_lightcone}, constructed in a ``top-down'' manner from a specific covariant theory \eqref{eq:S_HS_SD_GR}, works equally well in the more general setup $q^\alpha\bar q^{\dot\alpha}\del_{\alpha\dot\alpha}t\neq 0$. This can be attributed to the fact that we're working with a self-dual theory. Indeed, self-duality means that our equations depend on $q^\alpha$ but not on $\bar q^{\dot\alpha}$. As a result, the condition $q^\alpha\bar q^{\dot\alpha}\del_{\alpha\dot\alpha}t = 0$, or equivalently $q^\alpha\del_{\alpha\dot\alpha}t \sim \bar q_{\dot\alpha}$, simply never arises.

We will elaborate on these geometric comments in section \ref{sec:lightcone_covariance}. 

\subsubsection{Different scaling of the lightcone scalars under boosts} \label{sec:lightcone_ansatz:comparing:scaling}

For the moment, to further compare our lightcone formalism with that of \cite{Metsaev:2018xip,Skvortsov:2018uru}, let us place the two on similar footing by setting $q^\alpha\bar q^{\dot\alpha}\del_{\alpha\dot\alpha}t = 0$. We then note one further difference between the formalisms, in the behavior of the lightcone fields $\phi^{(s)},\psi^{(s)}$ under Lorentz boosts that rescale $(q^\alpha,\bar q^{\dot\alpha})\to (\rho q^\alpha,\rho\bar q^{\dot\alpha})$ while leaving $t$ unaffected. In the language of \cite{Metsaev:2018xip}, such boosts are enacted by the kinematical generator $J^{+-}$. In \cite{Metsaev:2018xip}, this operator acts on the lightcone fields strictly as a diffeomorphism, without any additional intrinsic rescaling. In contrast, in our formalism, it's easy to see from \eqref{eq:ansatz_2},\eqref{eq:psi} that $\phi^{(s)}$ and $\psi^{(s)}$ rescale intrinsically under $q^\alpha\to \rho q^\alpha$ as $\phi^{(s)}\to \rho^{-2s}\phi^{(s)}$ and $\psi^{(s)}\to \rho^{2s}\psi^{(s)}$ (the same is true in e.g. \cite{Siegel:1992wd}). This shows that our encoding of the physical degrees of freedom in terms of scalar fields is slightly different from the one in \cite{Metsaev:2018xip,Skvortsov:2018uru}. One can account for this difference by positing that our scalars and those of \cite{Metsaev:2018xip,Skvortsov:2018uru} are related via derivatives $(q^\alpha\bar q^{\dot\alpha}\del_{\alpha\dot\alpha})^s$ along the lightlike generators.

\section{Hidden Lorentz covariance of the $\Lambda\neq 0$ lightcone formalism} \label{sec:lightcone_covariance}

In this section, we take a look at the geometric meaning of the lightcone ansatz \eqref{eq:ansatz_1}-\eqref{eq:ansatz_2}, and of the lightcone formalism for $\Lambda\neq 0$ more generally. In particular, we will argue that the added complexity of the $\Lambda\neq 0$ setup carries a hidden virtue: the $\Lambda\neq 0$ lightcone formalism contains within itself the tools for changing the lightlike frame, and thus reclaiming Lorentz covariance. In section \ref{sec:lightcone_covariance:general}, we will detail this argument. Then, in section \ref{sec:lightcone_covariance:unfolding}, we'll discuss the surprising similarities between the resulting picture and Vasiliev's unfolded formalism. In section \ref{sec:lightcone_covariance:HS}, we'll connect the general discussion to the specifics of our ansatz \eqref{eq:ansatz_1}-\eqref{eq:ansatz_2} for HS self-dual GR.

\subsection{Changing the lightlike reference frame} \label{sec:lightcone_covariance:general}

Generally speaking, the lightcone formalism for field theory is a bargain with upsides and downsides. In particular, we get rid of gauge symmetry and focus on physical degrees of freedom, at the cost of sacrificing Lorentz covariance by singling out a special lightlike direction. In the original context of Minkowski spacetime (i.e. $\Lambda=0$), this is straightforwardly true. Working in lightlike coordinates:
\begin{align}
    ds^2 = 2dx^+ dx^- + dx^i dx^i \ , \label{eq:Minkowski_lightcone}
\end{align}
the special lightlike direction indices a foliation of spacetime into parallel null hyperplanes $x^+ = \const$; initial data for the field equations is then defined on one of these hyperplanes, e.g. $x^+=0$, and can be evolved to the next hyperplane through a first-order differential equation (note that the d'Alembertian $\Box = 2\del_+\del_- + \del_i\del_i$ is first-order in $x^+$). Since we cannot escape the chosen lightlike foliation, Lorentz covariance is lost.

It is instructive to consider this $\Lambda=0$ situation from the point of view of the lightlike conformal boundary $\scri$ of Minkowski space. Indeed, every null hyperplane in Minkowski is just the lightcone of a point on $\scri$. In particular, a foliation of Minkowski space into \emph{parallel} null hyperplanes consists of the lightcones of points that sit on a \emph{lightray} $\mathcal{R}\subset\scri$. Starting with initial data on a particular hyperplane, i.e. on the lightcone of a particular point $P\in\mathcal{R}$, we evolve onto the next hyperplane, i.e. onto the lightcone of the next point along $\mathcal{R}$. Again, Lorentz covariance is lost, since we can only evolve along a single lightlike direction: since $\scri$ is itself a null hypersurface, every origin point $P\in\scri$ belongs to a unique lightray $\mathcal{R}\subset\scri$, along which we must advance to the next origin point. 

Now, consider the AdS (i.e. $\Lambda<0$) lightcone formalism developed in \cite{Metsaev:1999ui,Metsaev:2003cu,Metsaev:2018xip}. There, we work with lightcone Poincare coordinates on AdS:
\begin{align}
    ds^2 = \frac{dz^2 + dx^+ dx^- + (dx^1)^2}{z^2} \ , \label{eq:AdS_lightcone}
\end{align}
and again consider the foliation into $x^+=\const$ lightlike hypersurfaces. In the curved AdS metric \eqref{eq:AdS_lightcone}, these are no longer flat hyperplanes. Nevertheless, just like in the Minkowski case, they are the lightcones of origin points $P$ that sit on a lightray $\mathcal{R}$ on the spacetime's conformal boundary. In the coordinates \eqref{eq:AdS_lightcone}, these points are doubly singular: in addition to being on the AdS boundary $z\to 0$, they are also at lightlike infinity with respect to the flat boundary coordinates $(x^+,x^-,x)$. However, this second singularity is a coordinate artifact -- in reality, all points $P$ and all lightrays $\mathcal{R}$ on the AdS boundary are equivalent. 

Now, the key point is that, in the AdS case, the boundary is not lightlike, but Lorentzian. As a result, the boundary point $P$ has not one, but many lightrays $\mathcal{R}$ passing through it. These correspond to different choices of Poincare coordinates \eqref{eq:AdS_lightcone}, each with its own lightlike direction $\del_+$ along which to evolve onto the next lightcone. In the terminology of \cite{Metsaev:2018xip}, these different choices are related via the ``kinematical'' subgroup of the AdS isometries (i.e. the subgroup that preserves $P$ and its lightcone) -- specifically, via the generator $K^1$. Thus, the greater complexity of the AdS lightcone coordinates \eqref{eq:AdS_lightcone} as opposed to the Minkowski ones \eqref{eq:Minkowski_lightcone} encodes genuine geometric content: while the Minkowski lightcone coordinates just choose a boundary point $P$, the AdS ones choose also a preferred boundary lightray $\mathcal{R}$ passing through it. This choice of a preferred lightlike direction at $P$ can be called a ``lightlike reference frame''.

We arrive at the following picture. In the Minkowski lightcone formalism, we are forced to evolve from (the null hyperplane originating at) $P$ in a predetermined direction, following the predetermined boundary lightray $\mathcal{R}$. In AdS, on the other hand, we can also apply kinematical generators to ``change the lightlike reference frame'', i.e. choose a new boundary lightray $\mathcal{R}$ along which to evolve. In this way, Lorentz covariance is essentially restored: by rotating our reference frame and advancing along different boundary lightrays, we can evolve our fields from the lightcone of $P$ to the lightcone of any other boundary point $P'$, without being bound by the original preferred lightlike direction $\del_+$.

This picture gets further upgraded if the lightcone formalism generalizes to non-orthogonal $\del_-$ and $\del_z$, as is the case for HS self-dual GR in section \ref{sec:lightcone_ansatz}. As discussed in section \ref{sec:lightcone_ansatz:comparing:bulk}, this places the origin point $P$ and the lightray $\mathcal{R}$ in the \emph{bulk}. The lightcone formalism's ``kinematical subgroup'' is then just the bulk Lorentz group at $P$, which can be used to set the ``lightlike reference frame'', i.e. the preferred lightray $\mathcal{R}$, to any lightlike direction at $P$. With this freedom, we can use the lightcone formalism to evolve our fields from the lightcone of $P$ to the lightcone of any other bulk point $P'$.

\subsection{Comparison with the unfolding formalism} \label{sec:lightcone_covariance:unfolding}

The above perspective on the $\Lambda\neq 0$ lightcone formalism is surprisingly similar to Vasiliev's \emph{unfolded} formalism for field theory. At first glance, these two formalisms are polar opposites: 
\begin{itemize}
  \item The lightcone formalism encodes initial data on a null hyperplane, while the unfolded formalism encodes it at a single point, via master fields such as $C(x^a;y^\alpha,\bar y^{\dot\alpha})$. 
  \item The lightcone formalism removes most of the components of a massless spin-$s$ field, leaving only the two ``physical'' ones (left-handed and right-handed). In contrast, the unfolded master fields contain not only all components, but \emph{every possible Lorentz representation} that encodes a given spin $s$. For instance, the 0-form master field $C(x;y,\bar y)$ contains all representations of the form $C_{\alpha_1\dots\alpha_{2s+k}\dot\alpha_1\dots\dot\alpha_k}(x) = C_{(\alpha_1\dots\alpha_{2s+k})(\dot\alpha_1\dots\dot\alpha_k)}(x)$ and $C_{\alpha_1\dots\alpha_k\dot\alpha_1\dots\dot\alpha_{2s+k}}(x) = C_{(\alpha_1\dots\alpha_k)(\dot\alpha_1\dots\dot\alpha_{2s+k})}(x)$.
\end{itemize}
On a closer look, though, these seemingly profound differences become rather superficial. As we discussed in section \ref{sec:lightcone_covariance:general}, a null hyperplane is nothing but the lightcone of a \emph{point} -- in the most general lightcone formalism, an ordinary bulk point. As for the infinite tower of Lorentz tensors $C_{\alpha_1\dots\alpha_{2s+k}\dot\alpha_1\dots\dot\alpha_k}(x),C_{\alpha_1\dots\alpha_k\dot\alpha_1\dots\dot\alpha_{2s+k}}(x)$ for each spin $s$, it was understood long ago by Penrose \cite{Penrose:1980yx} that these are nothing but the Taylor coefficients of the left-handed and right-handed field components \emph{along the lightcone of $x$} -- the very same components to which the spin-$s$ field is reduced in the lightcone formalism.

Another similarity is that in both formalisms, evolution is defined by a first-order differential equation. Indeed, in the lightcone formalism, the field equation is first-order in $\del_+$ (i.e. in derivatives that point towards the next null hyperplane or lightcone), whereas unfolded equations are always first-order in the exterior derivative $d$. The main difference is that unfolded equations evolve the master field in \emph{any} direction from the spacetime point $x$, whereas in the lightcone formalism, we can only move the lightcone's origin point along \emph{lightlike} directions. This is reflected in the fact that the unfolded formalism requires a 1-form master field $\Omega_a(x;y,\bar y)$ alongside the 0-form $C(x;y,\bar y)$, whereas the lightcone formalism requires a ``lightlike reference frame'' (encoded above in the choice of Poincare coordinates).

\subsection{Application to our lightcone ansatz for HS self-dual GR} \label{sec:lightcone_covariance:HS}

So far in this section, we've been discussing the $\Lambda\neq 0$ lightcone formalism somewhat abstractly. In particular, we talked about ``field data'' on lightcones, without much attention to the fields' tensor structure. Here, we will make some brief comments to fill this gap, in the context of our lightcone ansatz \eqref{eq:ansatz_1}-\eqref{eq:ansatz_2} for HS self-dual GR, which expresses the covariant connection $\omega_a(x^a,y^\alpha)$ in terms of the lightcone scalars $\phi(x^a,u)$. Once again, our comments will also apply to the pure spin-2 sector, i.e. to self-dual GR itself.

The first comment concerns what is meant by ``lightcone'': are we talking about lightcones in pure (A)dS, or in the perturbed geometry defined by the spin-2 connection $\omega_a^{\alpha\beta}$? To be concrete, let us write down explicitly the vielbein, inverse vielbein and metric \cite{Neiman:2023bkq} that correspond to $\omega_a^{\alpha\beta}$ in the ansatz \eqref{eq:ansatz_1}-\eqref{eq:ansatz_2}:
\begin{align}
    e_a^{\alpha\dot\alpha}(x) &= \frac{1}{t}\left(\sigma_a^{\alpha\dot\alpha} + \sigma_a^{\beta\dot\beta}\,q^\alpha q_\beta q^\gamma q^\delta\!
    \left(\frac{t}{2}\,\del_\gamma{}^{\dot\alpha}\del_{\delta\dot\beta}\phi^{(2)}(x) - \del_\gamma{}^{\dot\alpha}t\,\del_{\delta\dot\beta}\phi^{(2)}(x) \right) \right) \ ; \label{eq:vielbein} \\
    e^a_{\alpha\dot\alpha}(x) &=  t\left(\sigma^a_{\alpha\dot\alpha} - \sigma^a_{\beta\dot\beta}\,q_\alpha q^\beta q^\gamma q^\delta\!
    \left(\frac{t}{2}\,\del_{\gamma\dot\alpha}\del_\delta{}^{\dot\beta}\phi^{(2)}(x) - \del_{\gamma\dot\alpha}t\,\del_\delta{}^{\dot\beta}\phi^{(2)}(x) \right) \right) \ ; \label{eq:inverse_vielbein} \\
    g_{ab}(x) &= \frac{1}{t^2}\left(\eta_{ab} - \sigma_{(a}^{\alpha\dot\alpha}\sigma_{b)}^{\beta\dot\beta}\,q_\alpha q_\beta q^\gamma q^\delta\!
      \left(\frac{t}{2}\,\del_{\gamma\dot\alpha}\del_{\delta\dot\beta}\phi^{(2)}(x) - \del_{\gamma\dot\alpha}t\,\del_{\delta\dot\beta}\phi^{(2)}(x) \right) \right) \ . \label{eq:metric}
\end{align}
In each of these, the first term describes pure $dS_4$, while the second term is the deformation due to the right-handed spin-2 degree of freedom $\phi^{(2)}(x)$. Now, our observation is that the ``null hyperplanes'' (actually, lightcones) $q_\alpha\bar q_{\dot\alpha}x^{\alpha\dot\alpha} = \const$ remain largely unchanged by the deformation. The reason is that the deformation terms in \eqref{eq:vielbein}-\eqref{eq:metric} are always along $q_\alpha$. Specifically, the following properties remain unchanged:
\begin{itemize}
    \item The hypersurface $q_\alpha\bar q_{\dot\alpha}x^{\alpha\dot\alpha} = \const$ remains null.
    \item The vector $\ell^a \equiv t^2\sigma^a_{\alpha\dot\alpha} q^\alpha\bar q^{\dot\alpha}$ remains an affine lightlike generator of this hypersurface.
    \item The complex vector $m^a \equiv t\sigma^a_{\alpha\dot\alpha}q^\alpha\bar\chi^{\dot\alpha}$ (for any $\bar\chi^{\dot\alpha}$ not along $\bar q^{\dot\alpha}$) remains a left-handed null vector within the hypersurface, i.e. orthogonal to itself and to $\ell^a$.
\end{itemize}
Thus, we can equally well use the pure (A)dS metric $\eta_{ab}/t^2$ or the exact one \eqref{eq:metric} to define the lightcones on which our field data lives. 

Next, let's consider the different components and gauge conditions satisfied by the connection \eqref{eq:ansatz_1}-\eqref{eq:ansatz_2}. We observe the following:
\begin{enumerate}
    \item For higher spins $s>2$, the connection components satisfy $e^a_{\beta_1\dot\alpha}\omega^{\beta_1\beta_2\dots\beta_{2s-2}}_a = 0$, i.e. $e^a_{\alpha\dot\alpha}\omega^{\beta_1\dots\beta_{2s-2}}_a$ is totally symmetric in its $2s-1$ left-handed spinor indices (again, this is true regardless of whether we use the exact inverse vielbein \eqref{eq:inverse_vielbein}, or its pure (A)dS version $t\sigma^a_{\alpha\dot\alpha}$). As explained in \cite{Krasnov:2021nsq}, this fixes the gauge freedom of HS diffeomorphisms.
    \item The component $\ell^a\omega_a(x,y)$ remains undeformed from its pure (A)dS value. This fixes the internal gauge freedom within the null hypersurface.
    \item The component $m^a\omega_a(x,y)$ \emph{also} remains undeformed from its pure (A)dS value. This isn't a further gauge condition, but rather an expression of self-duality.
    \item The component $\bar m^a\omega_a(x,y)$ is non-trivial, and is the one that carries the field data on the lightcone.
    \item The component $n^a\omega_a(x,y)$, where $n^a\equiv \sigma^a_{\alpha\dot\alpha}\chi^\alpha\bar\chi^{\dot\alpha}$, is also non-trivial. However, since $n^a$ points outside our null hypersurface, this isn't further dynamical field data. Rather, it encodes the evolution of the internal gauge parameter from one null hypersurface to the next.
\end{enumerate}
Now, consider a change of lightlike reference frame in the sense of section \ref{sec:lightcone_covariance:general}. Namely, consider staying on the same lightcone hypersurface, but switching to different Poincare coordinates so as to change the preferred lightlike direction, i.e. the direction of evolution towards the ``next'' lightcone. This will change the basis vectors $(\ell^a,m^a,\bar m^a,n^a)$ at each point of our lightcone, but will not affect the key gauge properties 1-3 above:
\begin{enumerate}
    \item The HS-diffeomorphism-fixing condition $e^a_{\beta_1\dot\alpha}\omega^{\beta_1\beta_2\dots\beta_{2s-2}}_a = 0$ is Lorentz-invariant, and thus unaffected.
    \item The direction of the lightlike generator $\ell^a$ at each point of our lightcone is unaffected, though its scaling will change. Thus, the fixing of internal gauge symmetry via $\ell^a\omega_a(x,y)$ remains unaffected.
    \item The direction of the left-handed totally-null bivector $\ell^{[a}m^{b]}$ is again unaffected. Together with the non-deformation of $\ell^a\omega_a(x,y)$, this means that $m^a\omega_a(x,y)$ also remains undeformed.
\end{enumerate}
The upshot is that within a given lightcone, changing the ``lightlike reference frame'' does not require any gauge transformation. However:
\begin{enumerate}[resume]
    \item The encoding of the non-trivial initial data component $\bar m^a\omega_a(x,y)$ in terms of the scalars $\phi(x,u)$ will change, i.e. $\phi(x,u)$ will transform non-trivially under a change in the lightlike frame.
    \item The component $n^a\omega_a(x,y)$, which encodes the \emph{gradient} of the internal gauge between our lightcone and the ``next'' one, will also change. This makes sense, since changing the lightlike frame specifically affects the direction of evolution, i.e. the choice of which lightcone is ``next''.
\end{enumerate}
To sum up, the details of the lightcone \emph{ansatz} \eqref{eq:ansatz_1}-\eqref{eq:ansatz_2}, namely the values of $\phi(x,u)$, are sensitive to changing the lightlike frame, whereas the \emph{gauge choice} imposed on the lightcone by \eqref{eq:ansatz_1}-\eqref{eq:ansatz_2} is insensitive.

\section{Outlook} \label{sec:outlook}

In this paper, we explored HS self-dual GR with cosmological constant. We pointed out that the restriction to even spins leads to a standard 4d spacetime described by self-dual GR. On the other hand, we found a simple 6d-covariant formulation, of which the original 4d formulation is a gauge-fixing. Conversely, by gauge-fixing further, we found a lightcone ansatz for the covariant fields, which leads to a lightcone action with only cubic interactions. Finally, we made some conceptual observations about the geometric meaning of the lightcone ansatz with $\Lambda\neq 0$, noting its hidden Lorentz covariance and startling similarity to the unfolded formalism.

The paper can be fairly described as a collection of loose ends, which lends itself to many open questions. Can we find an application for the 6d-covariant formalism from section \ref{sec:6d}? The most striking aspect of that formalism is the extent to which the spacetime coordinates $x^a$ and spinor coordinates $y^\alpha$ are placed on an equal footing. Can this somehow extend into the full theory of HS gravity, with the left-handed spinor $y^\alpha$ joined by its right-handed counterpart $\bar y^{\dot\alpha}$? Presumably, such a picture would be non-local w.r.t. all of $(x^a,y^\alpha,\bar y^{\dot\alpha})$, just as the HS self-dual GR sector is \emph{local} w.r.t. both $x^a$ and $y^\alpha$. 

Turning to the lightcone ansatz of section \ref{sec:lightcone_ansatz}, it would be interesting to investigate its integrability (note that the spin-2 sector, i.e. self-dual GR with $\Lambda\neq 0$, was suggested to be integrable in \cite{Lipstein:2023pih}). It would also be interesting to try and extend our ansatz to cover the full self-dual sector of HS gravity. In this context, the key feature of our ansatz would be its cubic-exactness, i.e. the absence of quartic or higher-order interactions. The full self-dual HS theory in the lightcone formalism is known to be cubic-exact for $\Lambda=0$ \cite{Skvortsov:2018jea,Skvortsov:2020wtf}. Might the same be true for $\Lambda\neq 0$? The precise properties of our lightcone ansatz, especially the differences from Metsaev's lightcone formalism (section \ref{sec:lightcone_ansatz:comparing}), may be relevant to this question.

It would be nice to find an application for the picture of section \ref{sec:lightcone_covariance}, in which the $\Lambda\neq 0$ lightcone formalism allows us to evolve from any lightcone hypersurface to any other, by evolving along different lightlike directions. In particular, one could apply it to the problem of scattering in the de Sitter static patch \cite{David:2019mos,Albrychiewicz:2020ruh,Albrychiewicz:2021ndv,Neiman:2023bkq}, where the challenge is to evolve the fields from one lightcone (a de Sitter observer's past cosmological horizon) to another lightcone (the future horizon). It would also be interesting to clarify to what extent the picture of section \ref{sec:lightcone_covariance} relies on $\Lambda\neq 0$. Specifically, is there a variation on the $\Lambda\neq 0$ lightcone formalism that would apply to lightcones of \emph{bulk} points in Minkowski space, for which there is no single predetermined lightlike direction along which to evolve? 

Finally, it would be interesting to further explore the similarity between the $\Lambda\neq 0$ lightcone formalism and the unfolded language (section \ref{sec:lightcone_covariance:unfolding}). One difference between the two is that the unfolded formalism doesn't involve a choice of ``lightlike reference frame''. However, the full Vasiliev equations famously involve an extra pair of spinor coordinates $(z^\alpha,\bar z^{\dot\alpha})$, on top of those prescribed by unfolding. Might there be a relation between $(z^\alpha,\bar z^{\dot\alpha})$ and lightlike reference frames?

Overall, our hope is that HS self-dual GR may point towards useful perspectives on the more complicated full theory of HS gravity, and ultimately on its physics in de Sitter space. 

\section*{Acknowledgements}

I am grateful to Kirill Krasnov, Evgeny Skvortsov, Tung Tran, Julian Lang, Slava Lysov and Mirian Tsulaia for discussions. This work was supported by the Quantum Gravity Unit of the Okinawa Institute of Science and Technology Graduate University (OIST). The conceptual picture was finalized at the 5th Mons Workshop on Higher-Spin Gauge Theories.


\begin{thebibliography}{99}

\bibitem{Vasiliev:1990en}
M.~A.~Vasiliev,
``Consistent equation for interacting gauge fields of all spins in (3+1)-dimensions,''
Phys. Lett. B \textbf{243}, 378-382 (1990)
doi:10.1016/0370-2693(90)91400-6

\bibitem{Vasiliev:1995dn} 
M.~A.~Vasiliev,
``Higher spin gauge theories in four-dimensions, three-dimensions, and two-dimensions,''
Int.\ J.\ Mod.\ Phys.\ D {\bf 5}, 763 (1996)
[hep-th/9611024].

\bibitem{Vasiliev:1999ba} 
M.~A.~Vasiliev,
``Higher spin gauge theories: Star product and AdS space,''
In *Shifman, M.A. (ed.): The many faces of the superworld* 533-610
[hep-th/9910096].

\bibitem{Klebanov:2002ja} 
I.~R.~Klebanov and A.~M.~Polyakov,
``AdS dual of the critical O(N) vector model,''
Phys.\ Lett.\ B {\bf 550}, 213 (2002)
[hep-th/0210114].

\bibitem{Sezgin:2002rt}
E.~Sezgin and P.~Sundell,
``Massless higher spins and holography,''
Nucl. Phys. B \textbf{644}, 303-370 (2002)
[erratum: Nucl. Phys. B \textbf{660}, 403-403 (2003)]
doi:10.1016/S0550-3213(02)00739-3
[arXiv:hep-th/0205131 [hep-th]].

\bibitem{Sezgin:2003pt}
E.~Sezgin and P.~Sundell,
``Holography in 4D (super) higher spin theories and a test via cubic scalar couplings,''
JHEP \textbf{07}, 044 (2005)
doi:10.1088/1126-6708/2005/07/044
[arXiv:hep-th/0305040 [hep-th]].

\bibitem{Giombi:2012ms} 
S.~Giombi and X.~Yin,
``The Higher Spin/Vector Model Duality,''
J.\ Phys.\ A {\bf 46}, 214003 (2013)
doi:10.1088/1751-8113/46/21/214003
[arXiv:1208.4036 [hep-th]].

\bibitem{David:2020fea}
A.~David and Y.~Neiman,
``Bulk interactions and boundary dual of higher-spin-charged particles,''
JHEP \textbf{03}, 264 (2021)
doi:10.1007/JHEP03(2021)264
[arXiv:2009.02893 [hep-th]].

\bibitem{Lysov:2022zlw}
V.~Lysov and Y.~Neiman,
``Higher-spin gravity\textquoteright{}s \textquotedblleft{}string\textquotedblright{}: new gauge and proof of holographic duality for the linearized Didenko-Vasiliev solution,''
JHEP \textbf{10}, 054 (2022)
doi:10.1007/JHEP10(2022)054
[arXiv:2207.07507 [hep-th]].

\bibitem{Anninos:2011ui} 
D.~Anninos, T.~Hartman and A.~Strominger,
``Higher Spin Realization of the dS/CFT Correspondence,''
Class.\ Quant.\ Grav.\  {\bf 34}, no. 1, 015009 (2017)
doi:10.1088/1361-6382/34/1/015009
[arXiv:1108.5735 [hep-th]].

\bibitem{Fotopoulos:2010ay} 
A.~Fotopoulos and M.~Tsulaia,
``On the Tensionless Limit of String theory, Off - Shell Higher Spin Interaction Vertices and BCFW Recursion Relations,''
JHEP {\bf 1011}, 086 (2010)
doi:10.1007/JHEP11(2010)086
[arXiv:1009.0727 [hep-th]].

\bibitem{Taronna:2011kt}
M.~Taronna,
``Higher-Spin Interactions: four-point functions and beyond,''
JHEP \textbf{04}, 029 (2012)
doi:10.1007/JHEP04(2012)029
[arXiv:1107.5843 [hep-th]].

\bibitem{Vasiliev:2015wma}
M.~A.~Vasiliev,
``Star-Product Functions in Higher-Spin Theory and Locality,''
JHEP \textbf{06}, 031 (2015)
doi:10.1007/JHEP06(2015)031
[arXiv:1502.02271 [hep-th]].

\bibitem{Bekaert:2015tva} 
X.~Bekaert, J.~Erdmenger, D.~Ponomarev and C.~Sleight,
``Quartic AdS Interactions in Higher-Spin Gravity from Conformal Field Theory,''
JHEP {\bf 1511}, 149 (2015)
doi:10.1007/JHEP11(2015)149
[arXiv:1508.04292 [hep-th]].

\bibitem{Skvortsov:2015lja}
E.~D.~Skvortsov and M.~Taronna,
``On Locality, Holography and Unfolding,''
JHEP \textbf{11}, 044 (2015)
doi:10.1007/JHEP11(2015)044
[arXiv:1508.04764 [hep-th]].

\bibitem{Sleight:2017pcz} 
C.~Sleight and M.~Taronna,
``Higher-Spin Gauge Theories and Bulk Locality,''
Phys.\ Rev.\ Lett.\  {\bf 121}, no. 17, 171604 (2018)
doi:10.1103/PhysRevLett.121.171604
[arXiv:1704.07859 [hep-th]].

\bibitem{Ponomarev:2017qab}
D.~Ponomarev,
``A Note on (Non)-Locality in Holographic Higher Spin Theories,''
Universe \textbf{4}, no.1, 2 (2018)
doi:10.3390/universe4010002
[arXiv:1710.00403 [hep-th]].

\bibitem{Gelfond:2018vmi} 
O.~A.~Gelfond and M.~A.~Vasiliev,
``Homotopy Operators and Locality Theorems in Higher-Spin Equations,''
Phys.\ Lett.\ B {\bf 786}, 180 (2018)
doi:10.1016/j.physletb.2018.09.038
[arXiv:1805.11941 [hep-th]].

\bibitem{Didenko:2018fgx} 
V.~E.~Didenko, O.~A.~Gelfond, A.~V.~Korybut and M.~A.~Vasiliev,
``Homotopy Properties and Lower-Order Vertices in Higher-Spin Equations,''
J.\ Phys.\ A {\bf 51}, no. 46, 465202 (2018)
doi:10.1088/1751-8121/aae5e1
[arXiv:1807.00001 [hep-th]].

\bibitem{Didenko:2019xzz} 
V.~E.~Didenko, O.~A.~Gelfond, A.~V.~Korybut and M.~A.~Vasiliev,
``Limiting Shifted Homotopy in Higher-Spin Theory and Spin-Locality,''
JHEP {\bf 1912}, 086 (2019)
doi:10.1007/JHEP12(2019)086
[arXiv:1909.04876 [hep-th]].

\bibitem{Gelfond:2019tac}
O.~A.~Gelfond and M.~A.~Vasiliev,
``Spin-Locality of Higher-Spin Theories and Star-Product Functional Classes,''
JHEP \textbf{03}, 002 (2020)
doi:10.1007/JHEP03(2020)002
[arXiv:1910.00487 [hep-th]].

\bibitem{Vasiliev:2022med}
M.~A.~Vasiliev,
``Projectively-compact spinor vertices and space-time spin-locality in higher-spin theory,''
Phys. Lett. B \textbf{834}, 137401 (2022)
doi:10.1016/j.physletb.2022.137401
[arXiv:2208.02004 [hep-th]].

\bibitem{Neiman:2023orj}
Y.~Neiman,
``Quartic locality of higher-spin gravity in de Sitter and Euclidean anti-de Sitter space,''
Phys. Lett. B \textbf{843}, 138048 (2023)
doi:10.1016/j.physletb.2023.138048
[arXiv:2302.00852 [hep-th]].

\bibitem{Krasnov:2016emc}
K.~Krasnov,
``Self-Dual Gravity,''
Class. Quant. Grav. \textbf{34}, no.9, 095001 (2017)
doi:10.1088/1361-6382/aa65e5
[arXiv:1610.01457 [hep-th]].

\bibitem{Bardeen:1995gk}
W.~A.~Bardeen,
``Selfdual Yang-Mills theory, integrability and multiparton amplitudes,''
Prog. Theor. Phys. Suppl. \textbf{123}, 1-8 (1996)
doi:10.1143/PTPS.123.1

\bibitem{Rosly:1996vr}
A.~A.~Rosly and K.~G.~Selivanov,
``On amplitudes in selfdual sector of Yang-Mills theory,''
Phys. Lett. B \textbf{399}, 135-140 (1997)
doi:10.1016/S0370-2693(97)00268-2
[arXiv:hep-th/9611101 [hep-th]].

\bibitem{Mason:2009afn}
L.~J.~Mason and D.~Skinner,
``Gravity, Twistors and the MHV Formalism,''
Commun. Math. Phys. \textbf{294}, 827-862 (2010)
doi:10.1007/s00220-009-0972-4
[arXiv:0808.3907 [hep-th]].

\bibitem{David:2019mos}
A.~David, N.~Fischer and Y.~Neiman,
``Spinor-helicity variables for cosmological horizons in de Sitter space,''
Phys. Rev. D \textbf{100}, no.4, 045005 (2019)
doi:10.1103/PhysRevD.100.045005
[arXiv:1906.01058 [hep-th]].

\bibitem{Albrychiewicz:2020ruh}
E.~Albrychiewicz and Y.~Neiman,
``Scattering in the static patch of de Sitter space,''
Phys. Rev. D \textbf{103}, no.6, 065014 (2021)
doi:10.1103/PhysRevD.103.065014
[arXiv:2012.13584 [hep-th]].

\bibitem{Albrychiewicz:2021ndv}
E.~Albrychiewicz, Y.~Neiman and M.~Tsulaia,
``MHV amplitudes and BCFW recursion for Yang-Mills theory in the de Sitter static patch,''
JHEP \textbf{09}, 176 (2021)
doi:10.1007/JHEP09(2021)176
[arXiv:2105.07572 [hep-th]].

\bibitem{Neiman:2023bkq}
Y.~Neiman,
``Self-dual gravity in de Sitter space: Light-cone ansatz and static-patch scattering,''
Phys. Rev. D \textbf{109}, no.2, 024039 (2024)
doi:10.1103/PhysRevD.109.024039
[arXiv:2303.17866 [gr-qc]].

\bibitem{Ponomarev:2016lrm}
D.~Ponomarev and E.~D.~Skvortsov,
``Light-Front Higher-Spin Theories in Flat Space,''
J. Phys. A \textbf{50}, no.9, 095401 (2017)
doi:10.1088/1751-8121/aa56e7
[arXiv:1609.04655 [hep-th]].

\bibitem{Skvortsov:2018jea}
E.~D.~Skvortsov, T.~Tran and M.~Tsulaia,
``Quantum Chiral Higher Spin Gravity,''
Phys. Rev. Lett. \textbf{121}, no.3, 031601 (2018)
doi:10.1103/PhysRevLett.121.031601
[arXiv:1805.00048 [hep-th]].

\bibitem{Skvortsov:2020wtf}
E.~Skvortsov, T.~Tran and M.~Tsulaia,
``More on Quantum Chiral Higher Spin Gravity,''
Phys. Rev. D \textbf{101}, no.10, 106001 (2020)
doi:10.1103/PhysRevD.101.106001
[arXiv:2002.08487 [hep-th]].

\bibitem{Skvortsov:2018uru}
E.~Skvortsov,
``Light-Front Bootstrap for Chern-Simons Matter Theories,''
JHEP \textbf{06}, 058 (2019)
doi:10.1007/JHEP06(2019)058
[arXiv:1811.12333 [hep-th]].

\bibitem{Skvortsov:2022syz}
E.~Skvortsov and R.~Van Dongen,
``Minimal models of field theories: Chiral higher spin gravity,''
Phys. Rev. D \textbf{106}, no.4, 045006 (2022)
doi:10.1103/PhysRevD.106.045006
[arXiv:2204.10285 [hep-th]].

\bibitem{Sharapov:2022faa}
A.~Sharapov, A.~Sharapov, E.~Skvortsov, E.~Skvortsov, A.~Sukhanov, A.~Sukhanov, R.~Van Dongen and R.~Van Dongen,
``Minimal model of Chiral Higher Spin Gravity,''
JHEP \textbf{09}, 134 (2022)
[erratum: JHEP \textbf{02}, 183 (2023)]
doi:10.1007/JHEP09(2022)134
[arXiv:2205.07794 [hep-th]].

\bibitem{Sharapov:2022awp}
A.~Sharapov and E.~Skvortsov,
``Chiral higher spin gravity in (A)dS4 and secrets of Chern\textendash{}Simons matter theories,''
Nucl. Phys. B \textbf{985}, 115982 (2022)
doi:10.1016/j.nuclphysb.2022.115982
[arXiv:2205.15293 [hep-th]].

\bibitem{Didenko:2022qga}
V.~E.~Didenko,
``On holomorphic sector of higher-spin theory,''
JHEP \textbf{10}, 191 (2022)
doi:10.1007/JHEP10(2022)191
[arXiv:2209.01966 [hep-th]].

\bibitem{Krasnov:2011up}
K.~Krasnov,
``Gravity as a diffeomorphism invariant gauge theory,''
Phys. Rev. D \textbf{84}, 024034 (2011)
doi:10.1103/PhysRevD.84.024034
[arXiv:1101.4788 [hep-th]].

\bibitem{Krasnov:2011pp}
K.~Krasnov,
``Pure Connection Action Principle for General Relativity,''
Phys. Rev. Lett. \textbf{106}, 251103 (2011)
doi:10.1103/PhysRevLett.106.251103
[arXiv:1103.4498 [gr-qc]].

\bibitem{Capovilla:1989ac}
R.~Capovilla, T.~Jacobson and J.~Dell,
``General Relativity Without the Metric,''
Phys. Rev. Lett. \textbf{63}, 2325 (1989)
doi:10.1103/PhysRevLett.63.2325

\bibitem{Capovilla:1990qi}
R.~Capovilla, T.~Jacobson and J.~Dell,
``GRAVITATIONAL INSTANTONS AS SU(2) GAUGE FIELDS,''
Class. Quant. Grav. \textbf{7}, L1-L3 (1990)
doi:10.1088/0264-9381/7/1/001

\bibitem{Krasnov:2021nsq}
K.~Krasnov, E.~Skvortsov and T.~Tran,
``Actions for self-dual Higher Spin Gravities,''
JHEP \textbf{08}, 076 (2021)
doi:10.1007/JHEP08(2021)076
[arXiv:2105.12782 [hep-th]].

\bibitem{Herfray:2016qvg}
Y.~Herfray,
``Pure Connection Formulation, Twistors and the Chase for a Twistor Action for General Relativity,''
J. Math. Phys. \textbf{58}, no.11, 112505 (2017)
doi:10.1063/1.5012268
[arXiv:1610.02343 [hep-th]].

\bibitem{Plebanski:1975wn}
J.~F.~Plebanski,
``Some solutions of complex Einstein equations,''
J. Math. Phys. \textbf{16}, 2395-2402 (1975)
doi:10.1063/1.522505

\bibitem{Siegel:1992wd}
W.~Siegel,
``Selfdual N=8 supergravity as closed N=2 (N=4) strings,''
Phys. Rev. D \textbf{47}, 2504-2511 (1993)
doi:10.1103/PhysRevD.47.2504
[arXiv:hep-th/9207043 [hep-th]].

\bibitem{Metsaev:2005ar}
R.~R.~Metsaev,
``Cubic interaction vertices of massive and massless higher spin fields,''
Nucl. Phys. B \textbf{759}, 147-201 (2006)
doi:10.1016/j.nuclphysb.2006.10.002
[arXiv:hep-th/0512342 [hep-th]].

\bibitem{Metsaev:2018xip}
R.~R.~Metsaev,
``Light-cone gauge cubic interaction vertices for massless fields in AdS(4),''
Nucl. Phys. B \textbf{936}, 320-351 (2018)
doi:10.1016/j.nuclphysb.2018.09.021
[arXiv:1807.07542 [hep-th]].

\bibitem{Herfray:2022prf}
Y.~Herfray, K.~Krasnov and E.~Skvortsov,
``Higher-spin self-dual Yang-Mills and gravity from the twistor space,''
JHEP \textbf{01}, 158 (2023)
doi:10.1007/JHEP01(2023)158
[arXiv:2210.06209 [hep-th]].

\bibitem{Penrose:1976js}
R.~Penrose,
``Nonlinear Gravitons and Curved Twistor Theory,''
Gen. Rel. Grav. \textbf{7}, 31-52 (1976)
doi:10.1007/BF00762011

\bibitem{Ward:1980am}
R.~S.~Ward,
``Self-dual space-times with cosmological constant,''
Commun. Math. Phys. \textbf{78}, 1-17 (1980)
doi:10.1007/BF01941967

\bibitem{Metsaev:1999ui}
R.~R.~Metsaev,
``Light cone form of field dynamics in Anti-de Sitter space-time and AdS / CFT correspondence,''
Nucl. Phys. B \textbf{563}, 295-348 (1999)
doi:10.1016/S0550-3213(99)00554-4
[arXiv:hep-th/9906217 [hep-th]].

\bibitem{Metsaev:2003cu}
R.~R.~Metsaev,
``Massive totally symmetric fields in AdS(d),''
Phys. Lett. B \textbf{590}, 95-104 (2004)
doi:10.1016/j.physletb.2004.03.057
[arXiv:hep-th/0312297 [hep-th]].

\bibitem{Penrose:1980yx}
R.~Penrose,
``NULL HYPERSURFACE INITIAL DATA FOR CLASSICAL FIELDS OF ARBITRARY SPIN AND FOR GENERAL RELATIVITY,''
Gen. Rel. Grav. \textbf{12}, 225-264 (1980)
doi:10.1007/BF00756234

\bibitem{Lipstein:2023pih}
A.~Lipstein and S.~Nagy,
``Self-Dual Gravity and Color-Kinematics Duality in AdS4,''
Phys. Rev. Lett. \textbf{131}, no.8, 081501 (2023)
doi:10.1103/PhysRevLett.131.081501
[arXiv:2304.07141 [hep-th]].

\end{thebibliography}
\end{document}